\documentclass[pre,twocolumn,twoside,byrevtex,superscriptaddress,floatfix]{revtex4-1}

\newcommand{\plainlatexonly}[1]{}
\newcommand{\revtexlatexswitch}[2]{#1}

\usepackage{currfile}
\usepackage{color}

\usepackage{graphicx,epsfig,verbatim,enumerate}
\usepackage{amssymb,amsmath}
\usepackage{ifthen}

\usepackage{longtable}

\usepackage{mathtools}

\newboolean{twocolswitch}

\newcommand{\sindex}[1]{}
\newcommand{\nindex}[1]{}

\newcommand{\etal}{\textit{et al.}}
\newcommand{\www}[1]{\url{#1}}

\usepackage{lettrine}

\setboolean{twocolswitch}{true}

\begin{document}

\title{\protect
Social media appears to affect the timing, location, and severity of school shootings
}

\author{
  \firstname{Javier} 
  \surname{Garcia-Bernardo}
}
\email{javgarber@gmail.com}

\affiliation{CORPNET, University of Amsterdam, Nieuwe Achtergracht 166, 1018 WV,
Amsterdam, The Netherlands}

\thanks{To whom correspondence should be addressed}

\author{
  \firstname{Hong }
  \surname{Qi}
}

\affiliation{
  Department of Physics, University of Miami, Coral Gables, FL 33124,
  USA
}

\author{
  \firstname{James M.}
  \surname{Shultz}
}

\affiliation{
  Center for Disaster \& Extreme Event Preparedness (DEEP Center),
  University of Miami, Miller School of Medicine, Miami, FL 33124, USA
}

\author{
\firstname{Alyssa M.}
\surname{Cohen}
}

\affiliation{10858 Limeberry Drive, Cooper City, FL 33026, USA}

\author{
\firstname{Neil F.}
\surname{Johnson}
}
\email{njohnson@physics.miami.edu}

\affiliation{
  Department of Physics, University of Miami, Coral Gables, FL 33124,
  USA
}

\thanks{To whom correspondence should be addressed}

\author{
\firstname{Peter Sheridan}
\surname{Dodds}
}
\email{peter.dodds@uvm.edu}

\affiliation{Computational Story Lab,
  Vermont Complex Systems Center,
  Department of Mathematics \& Statistics,
  \& the Vermont Advanced Computing Core,
  University of Vermont,
  Burlington, VT 05401, USA}

\thanks{To whom correspondence should be addressed}

\date{\today}

\begin{abstract}
  \protect
  Over the past two decades, school shootings within the United States 
have repeatedly devastated communities and shaken public opinion.
Many of these attacks appear to be `lone wolf' ones driven by
specific individual motivations,
and the identification of precursor signals and hence
actionable policy measures would thus seem highly unlikely.
Here, we take a system-wide view and investigate 
the timing of school attacks and the dynamical
feedback with social media.
We identify a trend divergence in which college attacks have continued
to accelerate over the last 25 years 
while those carried out on K-12 schools have slowed down.
We establish the copycat effect in school shootings and uncover a
statistical association between social media chatter
and the probability of an attack in the following days. 
While hinting at causality, this relationship may also help mitigate the
frequency and intensity of future attacks.
 
\end{abstract}


\maketitle

\section*{Introduction}
Extensive research has been carried out on individual mass shooting
case studies, yielding a complex variety of causes revolving around
individual-centric factors such as mental illness, social rejection
and harassment~\cite{Newman2004,Flannery2013,Kimmel2003,Leary2003}
(see~\cite{Muschert2006} and~\cite{O2000} for reviews). 
A sociological model to understand and prevent attacks has been
proposed~\cite{Levin2009} and several solutions have been presented,
including community cohesion~\cite{Newman2004} and early-signals
detection~\cite{Wike2009, Borum2010}. 
Recent studies have analyzed the causes of school shootings at a larger scale, 
finding that school shootings can impact new attacks~\cite{towers2015a},
and that unemployment rates are correlated with shooting rates~\cite{Pah2017}.
Our work provides a significant advance on current understanding by
providing a collective level description beyond individual case
studies, accompanied by a rigorous mathematical framework. 
We establish the copycat effect in school shootings, 
providing evidence for the role that media plays in the spread of school shootings~\cite{Muschert2017}.
These results follow from our unified treatment of two complementary
databases (see Methods), the first (Shultz) of which includes fatals
attacks from 1990--March 2018, while the second (Everytown)
includes all incidents from 2013--March 2018, irrespective of
whether there were casualties. 
Furthermore, we include a database of mass killings (collected by USA
Today), covering attacks causing more than four casualties from
2006--July 2015, showing that the results are not exclusive of school
shootings, but consistent across high-profile types of violence. 

\begin{figure*}[htp!]
  \centering
  \includegraphics[width=0.9\textwidth]{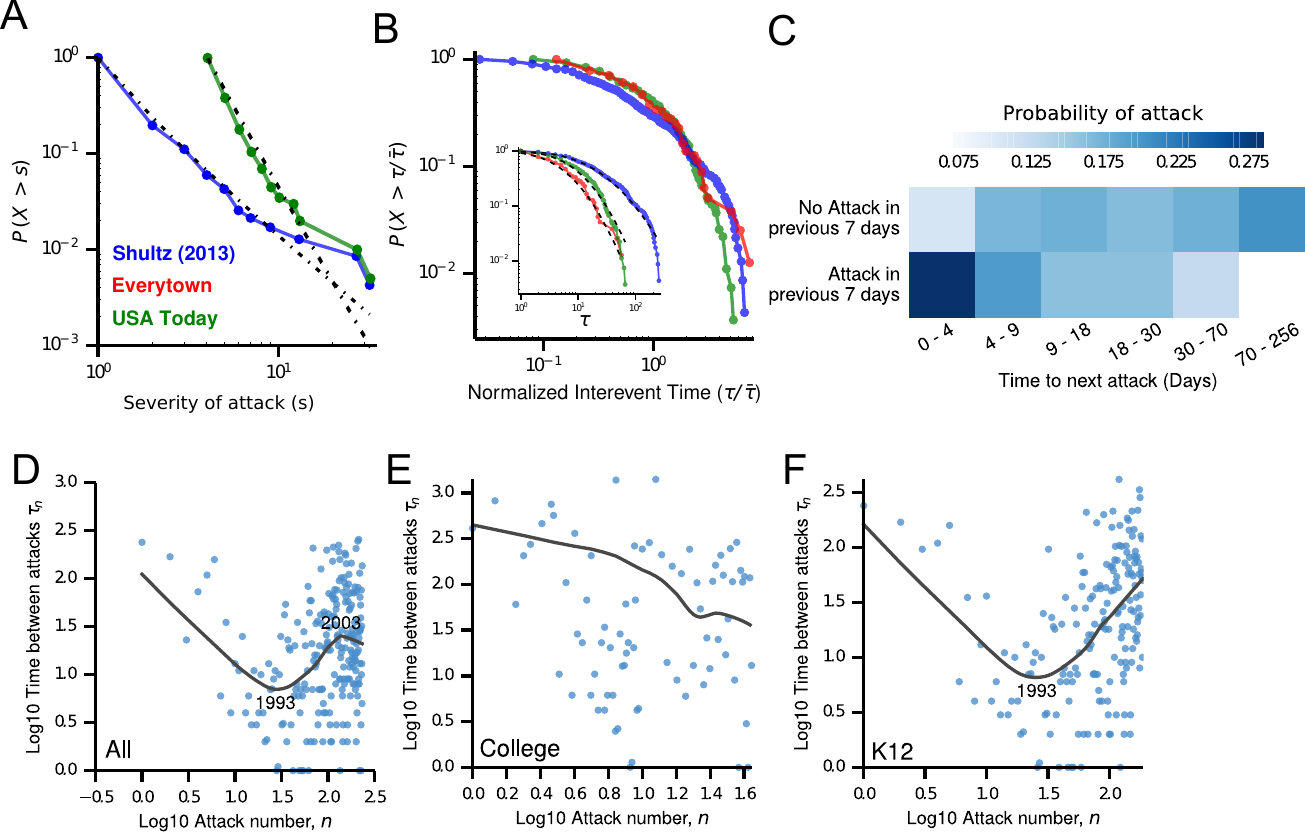}
  \caption{ 
    \textbf{Escalation patterns in school shootings.} 
    (A) Complementary Cumulative Distribution Function (CCDF) for
    event severity (dots and solid line) and best fit (dashed line) to
    power-law distribution. Note that the USA Today database only
    includes attacks with four or more victims.
    (B) CCDF for normalized interevent times (dots and solid
    line). Inset show the CCDF of the raw intereven times.
    (C) Probability of attack depending on the presence of an attack
    in the previous seven days. Each bin contains one sixth of the
    attacks.
    (D-F) The escalation plot, $\log_{10}{n}$ vs. $\log_{10}{\tau_n}$,
    for (D) \textit{All}, (E) \textit{K-12} and (F) \textit{College}
    attacks using the Schulz database (Methods).
    LOWESS fit ($\delta = 0$, $\alpha = 0.66$) is shown in dark gray,
    with the years where the trend changes annotated.
  }
  \label{fig:M_description}
\end{figure*}

\section*{Data characterization}
Many human activities have been shown to give rise to heavy-tail
distributions in the magnitude of the associated events and in the
interevent times.  Consistent with other human activities, we found
heavy-tail distributions in the attack size
(Fig. \ref{fig:M_description}A) and the timing of attacks
(Fig. \ref{fig:M_description}B) across the three databases studied.
Despite these data reflecting attacks with different characteristics,
all databases showed remarkable consistency in the interattack
distribution when normalized by the average waiting time
(Fig. \ref{fig:M_description}B).  Importantly, heavy-tail
distributions in the timing of attacks show a deviation from a random
Poisson process, where the event rate is uniform in time, and indicate
the presence of underlying factors.

The deviation from Poisson processes in complex systems has been
associated with burstiness~\cite{goh2008burstiness}, where events
cluster together in time
\revtexlatexswitch{(Figs. \ref{fig:S_clustering}A--D).}{Figs. S1A--D).}
Clustering can emerge from two mechanisms~\cite{goh2008burstiness}.
Firstly, it is related to the distribution of interevent times and can
be characterized by the normalized coefficient of variance $B =
\frac{(\sigma_{\tau}/\bar{\tau}) - 1}{(\sigma_{\tau}/\bar{\tau}) +
  1}$.  $B$ ranges between $-1$ for highly regular processes to $0$
for Poisson processes and $1$ for heavy-tail distributions.
Physiological complex systems such as hearbeats are highly regular,
while natural and human activities usually exhibit large burstiness
values~\cite{goh2008burstiness}.  Interestingly, the distribution of
time events is only moderately skewed, with $B = 0.142, 0.069
\textnormal{ and } -0.02$ for the Shultz, Everytown and USA Today
databases. These values contrast with the burstiness for other human
activities such as emailing, library loans, printing and calls, that
range between $0.2$ and $0.65$.  The second mechanism affecting
clustering is the memory of the system.  While natural activities
exhibit memory (e.g., large replicas follow large earthquakes), human
activities have low to no memory~\cite{goh2008burstiness}.  We
measured the memory of the system using autocorrelation, which ranges
between $-1$ for disassortative process---i.e., large (small)
interevent times follow small (large) interevent times, $0$ for no
correlation and $1$ for assortative processes.  In contrast to other
human activities, we found memory comparable to natural phenomena for
up to five attacks
\revtexlatexswitch{(Fig. \ref{fig:S_clustering}A).}{(Fig. S1A).}
Importantly, the
existance of memory is linked to a four-fold increase in the
probability of an attack in the days following a school shooting
(Fig. \ref{fig:M_description}C).  Given that the clustering not only
arises from a skewed distribution of the interevent times, but also
from memory, we hypothesize the existence of an external feedback loop
increasing the attack rate, that we later link to social media.

To further characterize the data, we analyzed the interevent time
distribution in detail.  We apply locally weighted scatterplot
smoothing (LOWESS) to the log-log plot of $\tau_n$ versus $n$
(Fig. \ref{fig:M_description}D--F), where the slope $b$ is an
indicator of changes in the attack rate~\cite{Johnson2013b}.  Figure
\ref{fig:M_description}D, containing all attacks, shows three regions
in time: From 1990 until 1993, the attack rate increased steadily
($b>0$).  From 1993 until 2003 there was a slowing down in the attacks
($b<0$), that was interrupted around 2004, when the escalation rate
again became positive.  Although the specific value of $b$ depends on
the correct determination of the first interevent time ($\tau_0$), the
results are robust to different values of $\tau_0$
\revtexlatexswitch{(Fig. \ref{fig:S_redblue_sensitivity}B--C).}{(Fig. S2B--C).}
The change in trend in
2004 shows that college attacks have been accelerating
(Fig. \ref{fig:M_description}E), while K-12 attacks have continued
slowing down (Fig. \ref{fig:M_description}F).  In the following
sections we describe and analyze the results of two models that have
been successfully applied to explain other forms of conflict: the
Hawkes process~\cite{hawkes1971a,hawkes1974cluster,laub2015a} and the
dynamical Red Queen or ``Red versus Blue" model~\cite{Johnson2013b}.

\begin{figure*}[htp!]
  \centering
  \includegraphics[width=0.9\textwidth]{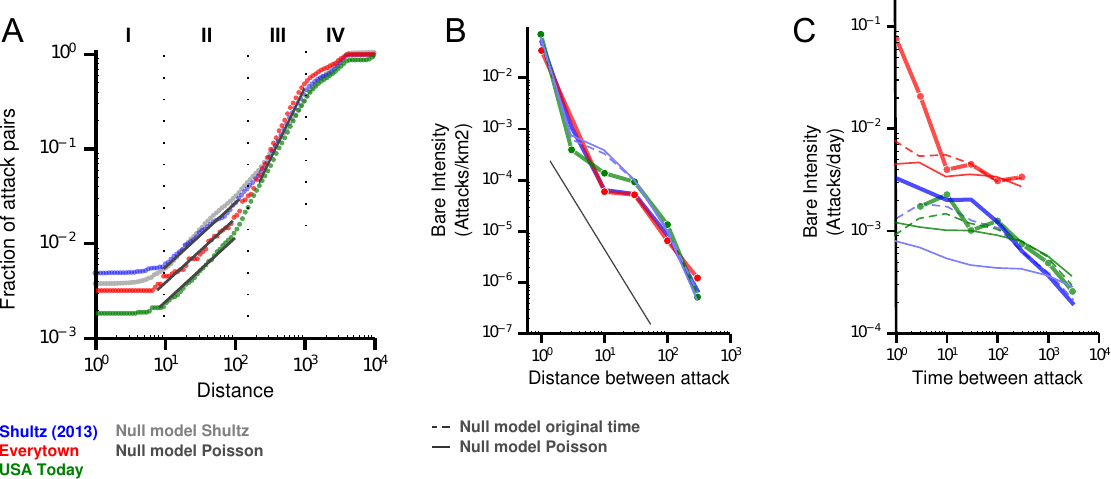}
  \caption{
    \textbf{Hawkes process model: Distance and time terms of the
      kernel function.} 
    (A) Fraction of attacks within a distance of each other for the Shultz and Everytown database and for the four null models, where the distribution of the population (drawn with probabily equal to
    the underlying US population or drawn at random from any coordinate in the United States) and the timing (equal to the observed time or drawn at random from a Poission distribution). The timing has only an effect in the kernel functions (B--C).
    (B) Intensity of attacks with respect to distance between
    attacks. A line with slope equal to 2 (i.e., the bare intensity
    decrease matching the increase in area) is shown for comparison.
    (C) Intensity of attacks with respect to time between attacks.
  }
  \label{fig:M_Hawkes}
\end{figure*}

\section*{Models}

\textit{Hawkes process}:
The Hawkes process is a self-exciting point process model described by
\begin{equation}
\lambda(x,t) = \mu + \sum_{i:t_i \le t}{g(x-x_i,t-t_i)},
\end{equation}\label{eq:hawkes}
where $\lambda(x,t)$ is the attack rate at position $x$ and
time $t$, $\mu$ is the background Poisson rate and $g(x-x_i,t-t_i)$ is
the contribution of the attack $i$ occurring at $x_i,t_i$. 
Hawkes processes have been typically used to study
earthquakes~\cite{ogata1988statistical}. 
In the case of seismicity, a triggered earthquake is followed by
aftershocks, which in turn activate new aftershocks creating a cascade
of events. 
This is modelled by separating earthquakes into background and
aftershocks, where background events occur with a specific background
rate and the probability of the aftershocks depends on the time and
distance from previous earthquakes according to the kernel $g(x,t)$. 
The kernel can be explicitly defined~\cite{ogata1988statistical} or
calculated using non-parametric methods~\cite{marsan2008extending}. 
The same modelling has also recently been successfully applied to
social phenomena, such as finance~\cite{hardiman2013critical},
crime~\cite{mohler2012self}, terrorism in
Irak~\cite{lewis2012self} and mass killings~\cite{towers2015a}. 
Here, we used the non-parametric method
from~\cite{marsan2008extending} to estimate the kernel $g(x,t)$ and
understand the mechanism by which school shootings trigger cascades of
attacks. 

Marsan and Lenglin{\'e}'s method~\cite{marsan2008extending} uses an
expectation-maximization algorithm on the binned events (earthquakes
in their case). 
It iteratively decouples the events into background and triggering
events using $\mu$ and $g$, and updates $\mu$ and $g$ using the new
decoupling of events until convergence is obtained. 
The original algorithm revealed a linear scaling between the magnitude
of the event and the probability of an aftershock. 
However, this either does not apply for school shootings or the
difference is too small to quantify given the sparsity of our data
\revtexlatexswitch{(Fig. \ref{fig:S_magnitudeSparse}A--B).}{(Fig. S3A--B).}
Therefore, we excluded the magnitude of the attack from the study and
calculate the relationship between the probability of new attacks
given the time and distance since previous attacks. 
In order to put our results in perspective, we created four null models
where we varied the distance and the timing between attacks.
In the first two models (I,II) the attacks were drawn at random from US cities with probability
proportional to their population (using the Geonames database). 
In the last two models (III,IV) the attacks were drawn at random from any coordinate in the United States.
Models I and III use a timing between attacks equal to the (Schulz or Everytown),
whereas in models II and IV the attacks occur as a Poisson process with $\lambda$
equal to the mean interevent time database ($\lambda_{Shultz} =
\overline{\tau_{Shultz}} = 39$ days and $\lambda_{Everytown} =
\overline{\tau_{Everytown}} = 6.5$ days). 

First, we analyze the the fraction of pairs of attacks that are
located within a specific distance of each other
(Fig. \ref{fig:M_Hawkes}). 
The distance between all attack pairs is similar to that expected if
the attacks were distributed proportional to the US population. 
Next, we analyzed the effect of time and distance in the spreading of
attacks. 
Figures \ref{fig:M_Hawkes}B--C show the two components of the kernel
function $g$, the intensity decrease as a function of the distance
between attacks (Fig. \ref{fig:M_Hawkes}B) and the decrease as a
function of time between attacks (Fig. \ref{fig:M_Hawkes}C). 
If the attacks were uniformly distributed, the algorithm would assign
a low weight to the kernel function.
However, we obtained a consistent form of the kernel function for all
databases studied. 
Both the distance between attacks and the time between attacks
diminish the probability of new attacks as an approximate power-law. 
Moreover, although the consistency in distance can be explained by the
underlying distribution of population (Fig. \ref{fig:M_Hawkes}B), the
consistency in timing cannot be explained by an underlying Poisson
process (Fig. \ref{fig:M_Hawkes}C).
Thus our results indicate that while the attacks occur approximately
at random in space, with the exception of within-town attacks, the
attacks do affect the timing of new shootings, increasing the rate of
attacks by a 3-10 fold, especially in the week following the attack.
Our results are consistent with those of Towers \etal~\cite{towers2015a}, 
where the kernel of the Hawkes process was explicitly modeled with a exponential decay on time.
While Towers \etal\ determined that an attack had a half-life of 8.9 days ([3.7-36.9] days) in subsequent attacks,
we estimate it on 7.3 days ([5.2 - 11.3] days).
The Hawkes model confirms the presence of attack cascades, and
quantifies the effect of distance and time in the probability of new
attacks.

\begin{figure*}[htp!]
  \centering
  \includegraphics[width=0.9\textwidth]{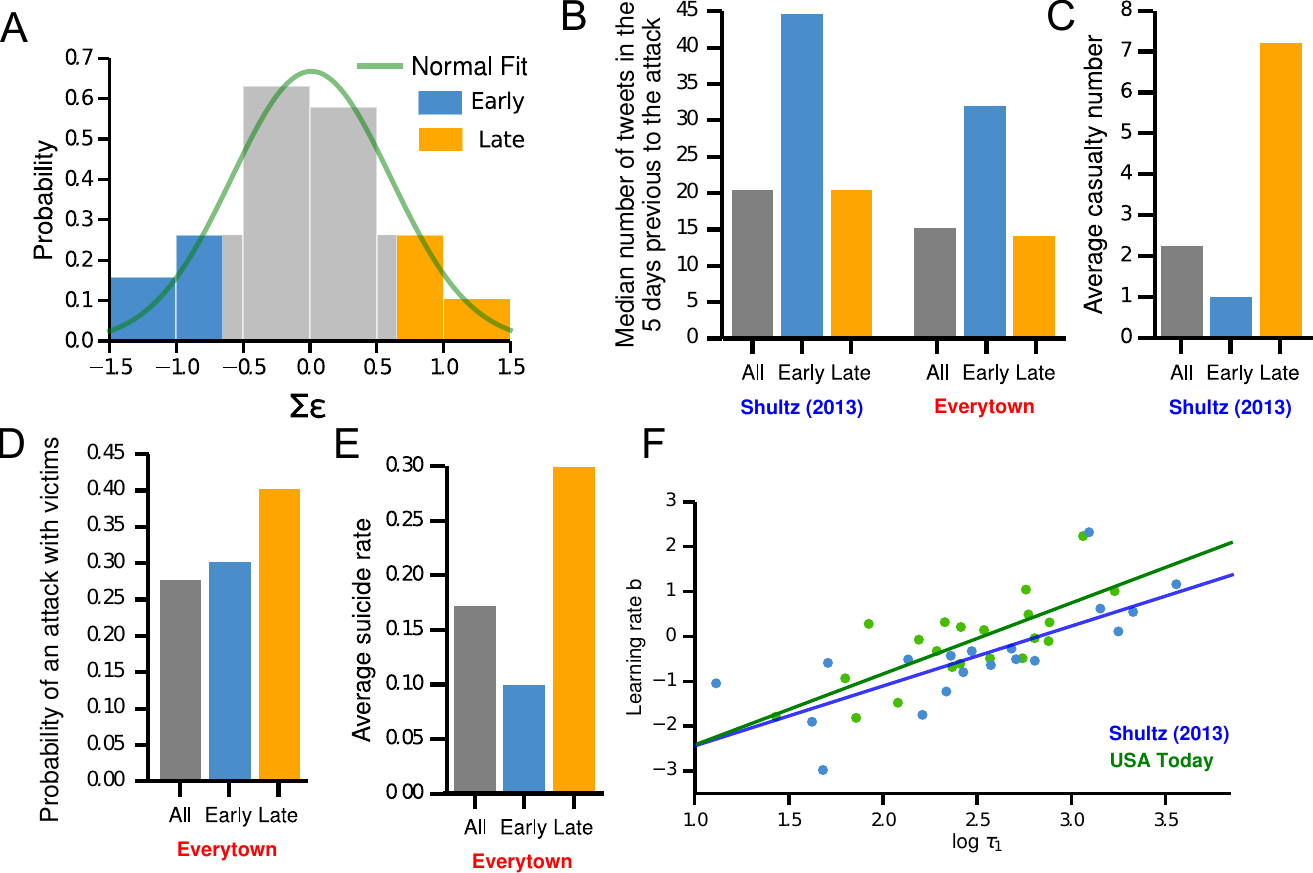}
  \caption{
    \textbf{Red vs Blue model: Attack characteristics.} 
    (A) Histogram of $\sum{\epsilon_j}$. \textit{Early} and
    \textit{Late} attacks are marked in blue and orange respectively.         
    (B--E) Attack characteristics for All (Grey), \textit{Early}
    (Blue) and \textit{Late} (Orange) attacks. (B) Number of tweets
    preceding the attacks. (C) Average casualty number. (D) Fraction
    of attacks with victims. (E) Fraction of attack ending in
    suicide. (F) Prediction plot, $\log_{10}{\tau_1}$ vs. $b$. All states with at least three events are considered. States above the $b=0$ line experienced an escalation in the number of attacks. 
The updated Shultz et al. database (until 2015, where Twitter data is available) was used for all plots except otherwise noted.
  }
  \label{fig:M_predict}
\end{figure*}

\textit{Red versus Blue model}: 
Empirical and theoretical studies have
shown that the trend in timings and distribution of severities of
attacks in human conflicts are described by the power laws $\tau_n =
\tau_1 n^{-b}$ and $p(s) \propto s^\alpha$ respectively, where
$\tau_n$ is the time between attacks $n$ and $n+1$, $b$ is the
escalation rate, $s$ is the attack severity, and $\alpha \simeq
2.5$~\cite{Johnson2011a,Johnson2013b}.
Positive values of the escalation rate $b$ reflect an increase in the
frequency of attacks with time, while the attack rate decreases if $b$
is negative. 
An explanatory model emerges from consideration of the confrontation
dynamics between two opponents~\cite{Johnson2011a}. 
In our case, the two `opponents' are the pool of potential attackers
which we call Red, none of whom are necessarily in contact with or
know each other, and Society which we call Blue. 
At any one instance, Red tends to hold a collective advantage $R$ over
Blue in that Red is largely an unknown threat group residing within
Blue. 
The size of this advantage depends on the number of potential
attackers and their resources. Each attack can affect the balance
between Red and Blue, for example by increasing
$R$~\cite{Johnson2011a,Johnson2013b}. 
It is reasonable to assume that the main changes in Red's lead $R$
over Blue occur just after a new attack, e.g., due to media
coverage. This is confirmed empirically by the increased probability
of a subsequent attack (Fig. \ref{fig:M_description}C), as well from
the results of the Hawkes model (Fig. \ref{fig:M_Hawkes}C). 
If the changes in $R$ are independent and identically distributed, the
Central Limit Theorem states that the typical value of $R$ after $n$
attacks, $R(n)$, will be proportional to $n^b$, where $b =
0.5$~\cite{Rudnick2010}. For the more general case where changes in
$R$ depend on the history of previous changes, $b$ will deviate from
$0.5$ corresponding to `anomalous'
diffusion~\cite{Klafter1987}. Taking the frequency of the attacks to
be proportional to Red's advantage over Blue, we obtain $\tau_n =
\tau_1 n^{-b}$. 

\begin{figure*}[htp!]
  \centering
  \includegraphics[width=0.9\textwidth]{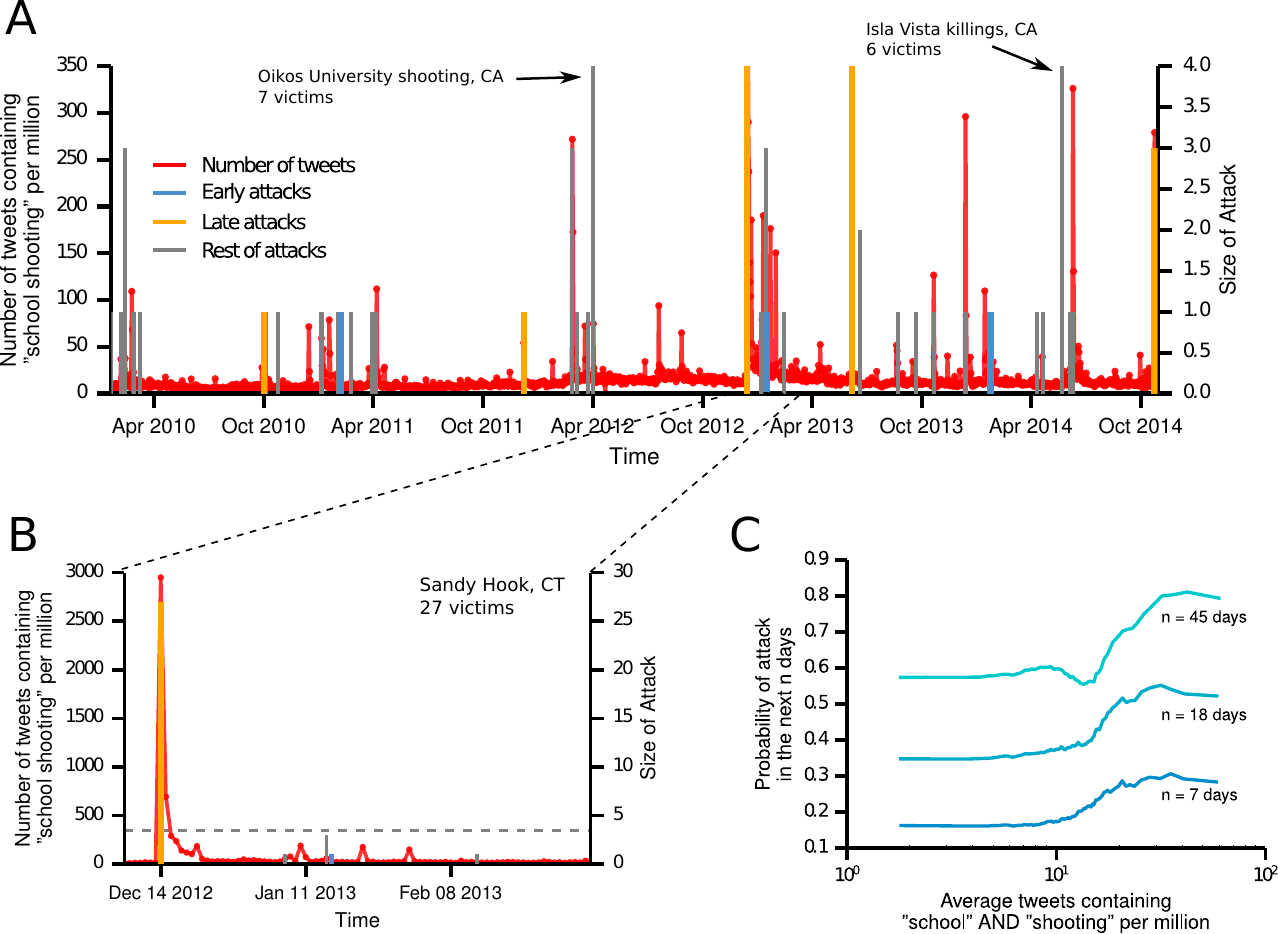}
  \caption{
    \textbf{Feedback loop between school shootings and mass
      media.} 
    (A) Time series of the number of tweets containing ``school" and
    ``shooting" (Red lines, left axis), and the severity of attacks
    (right axis) for \textit{Early} attacks (Blue), \textit{Late}
    attacks (Orange) and the rest (Grey). 
    (B) Sandy Hook incident. 
    (C) Probability of an attack happening in the 7, 18 or 45 days
    following attack $n$, as a function of the mean number of tweets
    with the words ``school" and ``shooting" at days $n$ and
    ${n+1}$. The Shultz database was used for all
    plots.
  } 
  \label{fig:M_twitter}
\end{figure*}

Our theory predicts that the time to the $n^{th}$ attack is determined
by the progress curve $\tau_n = \tau_1 n^{-b}$.
The progress curve assumes that the time to the next attack is
deterministic. However, in reality one can imagine that a series of
$N$ background processes would need to `fall into place' before a
potential attacker finds himself in an operational position to carry
out an attack and hence provide the $(n+1)^{th}$ attack. The
triggering of each of these $N$ processes may independently fluctuate
and so delay or accelerate the next attack. Similar to multiplicative
degradation processes in engineering, we assume that each of these
steps multiplies the expected time interval by a factor $(1+
\epsilon_j)$ where the stochastic variables $\epsilon_j$'s mimic these
exogenous factors. 
It is reasonable to assume that the values of the $\epsilon_j$'s are
independent and identically distributed, which means that their sum
(i.e., the noise term in the progress curve fit) is approximately
Gaussian distributed with zero mean (Fig. \ref{fig:M_predict}A). 
The observed time interval is now given by $\tau_n = \tau_1 n^b
\prod^N_{j=1}{(1+\epsilon_j)}$. 
It then follows that $\log{\tau_n} = \log{\tau_1} -b \log{n} +
\sum{\epsilon_j}$, since $\log{(1+\epsilon_j)} \simeq \epsilon_j$ if
$\epsilon_j \ll 1$. Hence the progress curve represents a straight
line fit through a maximum likelihood approach on a log-log plot,
exactly as assumed by our LOWESS analysis
(Fig. \ref{fig:M_description}D--F) where residuals are Gaussian
distributed. 
The attacks whose $\sum{\epsilon_j}$ deviates from zero are likely to
have distinctive characteristics. 
We labeled the attacks where $\sum{\epsilon_j}$ is larger than one
standard deviation as \textit{Late} and the ones where it is smaller
than one negative standard deviation as \textit{Early}
(Fig. \ref{fig:M_predict}A). 
We found that \textit{Early} attacks are correlated with high media
activity (Fig. \ref{fig:M_predict}B), as expected since those attacks
take place while the news about the previous one have not fade out. 
We also observed that \textit{Late} attacks are both more deadly
(Fig. \ref{fig:M_predict}C--D) and result more frequently in the
suicide of the attacker than \textit{Early} attacks
(Fig. \ref{fig:M_predict}E). 
We identify \textit{Late} attacks with planned attacks whose attackers
provide a continued leakage of clues over time~\cite{O2000}.

The \textit{Red versus Blue} model uncovers an unexpected
inter-relationship between the patterns of lone-wolf school attacks in
different geographical locations. 
If events in different locations were independent, one would not
expect any relationship between the $\log{\tau_1}$ and $b$ in
different locations. 
However Figure \ref{fig:M_predict}F shows that the opposite is true. 
The presence of a linear relationship among these different states, as
well as the presence of a significant kernel function in the
\textit{Hawkes} model, indicates that there is a common dynamical
factor influencing otherwise independent attackers across different
states. 
Our analysis suggests that the cause of this common dynamical factor
lies in modern media sources. 

\section*{The copycat effect}

Our hypothesis that the interaction between attacks is indirect
through the media is a phenomenon commonly known as the copycat
effect~\cite{Coleman2004}. 
This interaction can be attributed to an acute `issue-attention
cycle'~\cite{downs1972up} with the media reacting strongly to every
attack~\cite{Rocque2012}. 
Although the effect of mass media has been studied, evidence of
copycats has been anecdotal~\cite{O2000}. 
To analyze the role of social media (which echos and amplifies all
media), we obtained 72 million tweets containing the word ``shooting". 
From these, over 1.1 million tweets contained the word ``school". 
Figures \ref{fig:M_twitter}A--B visualizes the relationship between
the number of tweets containing the words ``school" and ``shooting"
with the \textit{Early} and \textit{Late} attacks. 
As expected given that a peak in Twitter activity follows every
attack, \textit{Early} attacks are correlated with periods of high
Twitter activity. 
To study the interaction between social media and school shootings, we
plotted the average number of tweets containing the words ``school"
and ``shooting" against the probability of an attack in the next 7, 17
and 44 days, corresponding to percentiles $25^{th}$, $50^{th}$, and
$75^{th}$ of the distribution of the days between attacks. 
Fig.~\ref{fig:M_twitter}C shows that the probability of an attack
increases with the number of tweets talking about school shootings. 
For example, the probability of an attack in the next week doubles
when the number of school shooting tweets increases from 10 to 50
tweets/million.
By contrast, tweets containing only ``shooting" or ``mass" and
``murder" did not show a pronounced effect
\revtexlatexswitch{(Fig.~\ref{fig:S_copycat_otherWords}).}{(Fig. S5).}
Our analysis thus confirms that social media publicity about school
shootings correlates with an increase in the probability of new
attacks.

\section*{Concluding remarks}

Our treatment goes towards explaining and
predicting the probablistic
escalation patterns in school shootings. 
Our theory is supported by analysis of an FBI dataset of active
shooting~\cite{FBI}
\revtexlatexswitch{(Figs. \ref{fig:S_FBIa} and \ref{fig:S_FBIb},}{(Figs. S6 and S7,}
Supplementary Information), and has implications in attack prevention
and mitigation. 
First, the discovery of distinct trends for college and K-12 attacks
should motivate policy makers to focus policy efforts in distinct ways
for these two educational settings. 
Second, the presence of underlying patterns in the data can improve
both short-term and long-term prediction of future trends, for example
by focusing the efforts in the cities where there has already been
already an attack. 
Finally, our analysis proves for the first time the copycat effect in
school shootings, a topic which has been analyzed primarily in a
narrative, case-by-case way to date. 
Our results do not contradict the fact that the psychological aspect
of the attacker is a key factor in an individual attack, or that
traditional prevention methods work, but instead draw a new collective
example of human conflict in which a small, dynamical, violent sector
of society confronts the remainder fueled by Blue's own informational
product (media). 

\section*{Methods}
\section*{Methods}

\subsection*{Databases}
We studied the following datasets:
\textbf{Everytown:} The attacks, with and without victims, were extracted from \url{http://everytown.org/}, containing all incidents from the period January 2013 to November 2014.
\textbf{Shultz:} The database for the period 1990--2013 gathered by Shultz et al.~\cite{Shultz2013} was updated with the Everytown database to include recent attacks with victims up to November 2014.
\textbf{USA Today:} The database for the period 2006--July 2015 gathered by \url{http://www.usatoday.com/},  including all attacks with four or more victims.
\textbf{Active shootings:} The date, size, age of the attacker and suicide result was obtained from the 2014 FBI report \textit{A Study of Active Shooter Incidents, 2000--2013}~\cite{FBI}.
\textbf{Twitter:} 57 billions tweets were analyzed in the period 2010 to November 2014, extracting over 72 million tweets with the word ``shooting", 1.1 million with the words ``shooting" and ``school", and 233 thousand with the words ``mass" and ``murder". 
 
\subsection*{Active shootings}
We repeated the analysis with the 160 active shootings events from the FBI database~\cite{FBI}. 
In this case, the distribution of attack sizes does not follow a power law (Fig. \ref{fig:S_FBIa}A). However, this is likely due to the definition of active shooting, where attacks with a low number of casualties do not tend to be included in the study. In agreement with the results of the report~\cite{FBI}, we find a steady rise in the frequency of attacks (Fig. \ref{fig:S_FBIa}B). 
Consistent with our results of school shootings, the time between the two first attacks is a good indicator of the subsequent escalation pattern (Fig. \ref{fig:S_FBIa}C). 
We found an interaction between attacks (Fig. \ref{fig:S_FBIa}D), which can be attributed to the copycat effect, since the probability of an attack in the subsequent 8, 19 and 35 days is correlated with the number of tweets containing ``shooting" (Fig. \ref{fig:S_FBIa}E), or ``school" and ``shooting" (Fig. \ref{fig:S_FBIa}F), but not ``mass" and ``murder" (Fig. \ref{fig:S_FBIa}G). 
We can define again \textit{Early} and \textit{Late} attacks (Fig. \ref{fig:S_FBIb}A), that correlate with Twitter activity (Fig. \ref{fig:S_FBIb}B--C). 
However, the size of the attacks in this case is not different for \textit{Early} and \textit{Late} attacks (Fig. \ref{fig:S_FBIb}D). 

Finally, we analyzed the correlation between age, size, and suicide rates (Fig. \ref{fig:S_FBIb}E--G). We found a positive correlation between age and attack size (Fig. \ref{fig:S_FBIb}E). 
Teenagers (ages 12--18) correlate with small size events (Fig. \ref{fig:S_FBIb}E) and low suicide rates (Fig. \ref{fig:S_FBIb}F). Young attackers (ages 18--38) exhibit high suicide rates (Fig. \ref{fig:S_FBIb}F). The size of the attack is not well correlated with suicide rates, with the exception of attacks without victims (Fig. \ref{fig:S_FBIb}G). 

\acknowledgments
We are grateful for funding from the Vermont Complex Systems Center
and use of the Vermont Advanced Computing Core.
PSD was supported by NSF CAREER Grant No. 0846668.

\clearpage

\newwrite\tempfile
\immediate\openout\tempfile=startsupp.txt
\immediate\write\tempfile{\thepage}
\immediate\closeout\tempfile

\setcounter{page}{1}
\renewcommand{\thepage}{S\arabic{page}}
\renewcommand{\thefigure}{S\arabic{figure}}
\renewcommand{\thetable}{S\arabic{table}}
\setcounter{figure}{0}
\setcounter{table}{0}

\textbf{Supporting Information}

\begin{figure*}[htp!]
  \centering
  \includegraphics[width=0.9\textwidth]{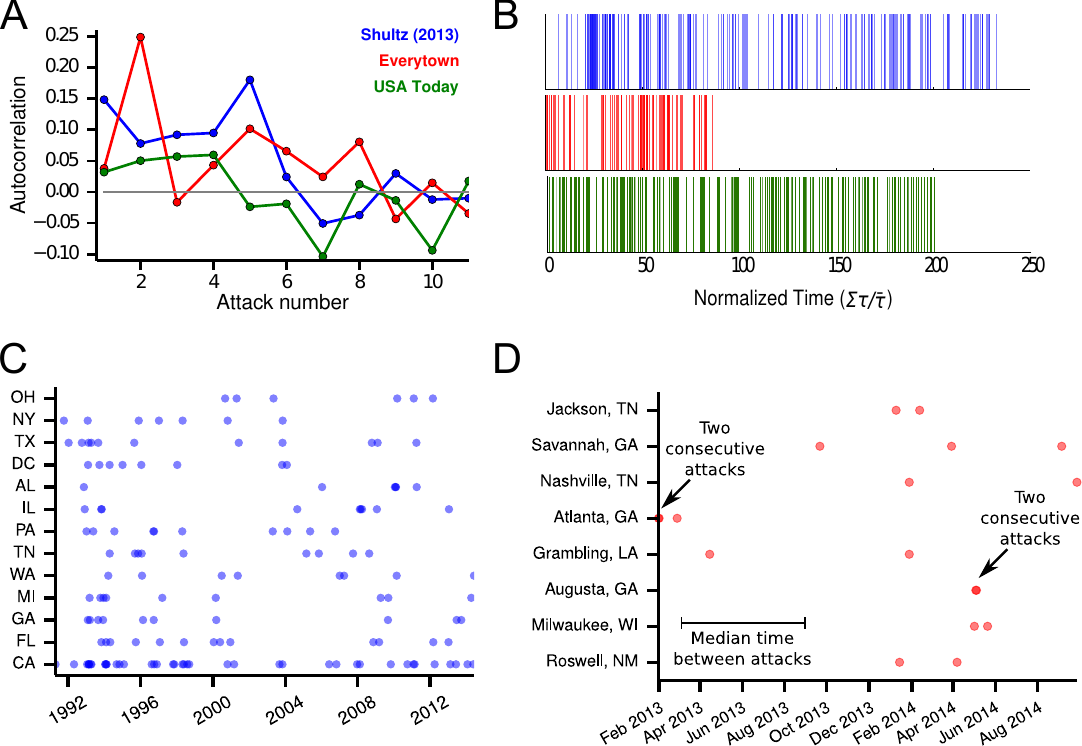}
  \caption{       
    (A) Autocorrelation ($AC_f(n) = \frac{1}{N-n}
    \sum_{t=0}^{N-n-1}{f(t+n)f(t)}$) for the interevent time series
    ($f$) for the Shultz, Everytown and USA Today databases at
    different lags ($n$). The interevent time series has been
    normalized by substracting the mean and dividing by the standard
    deviation.
    (B) Attack series using the normalized interevent time. Vertical
    bars correspond to individual attacks.
    (C) Attack series by state in the Shultz dataset.
    (D) Attack series in the 8 towns with two or more attacks in the Everytown dataset.
  }
  \label{fig:S_clustering}
\end{figure*}
        
\begin{figure*}[htp!]
  \centering
  \includegraphics[width=0.495\textwidth]{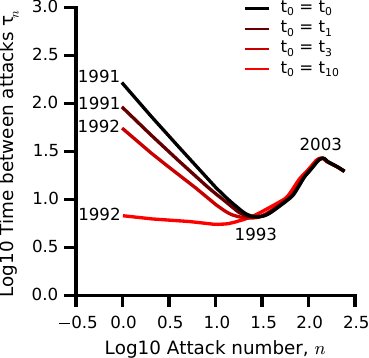}
  \caption{
    The progress plot $\log_{10}{n}$ vs. $\log_{10}{\tau_n}$,
    using all attacks ($[0-n]$), attacks $[1-n]$, $[3-n]$ and
    $[10-n]$ in the Shultz database.
  } 
  \label{fig:S_redblue_sensitivity}
\end{figure*}

\begin{figure*}[htp!]
  \centering
  \includegraphics[width=0.495\textwidth]{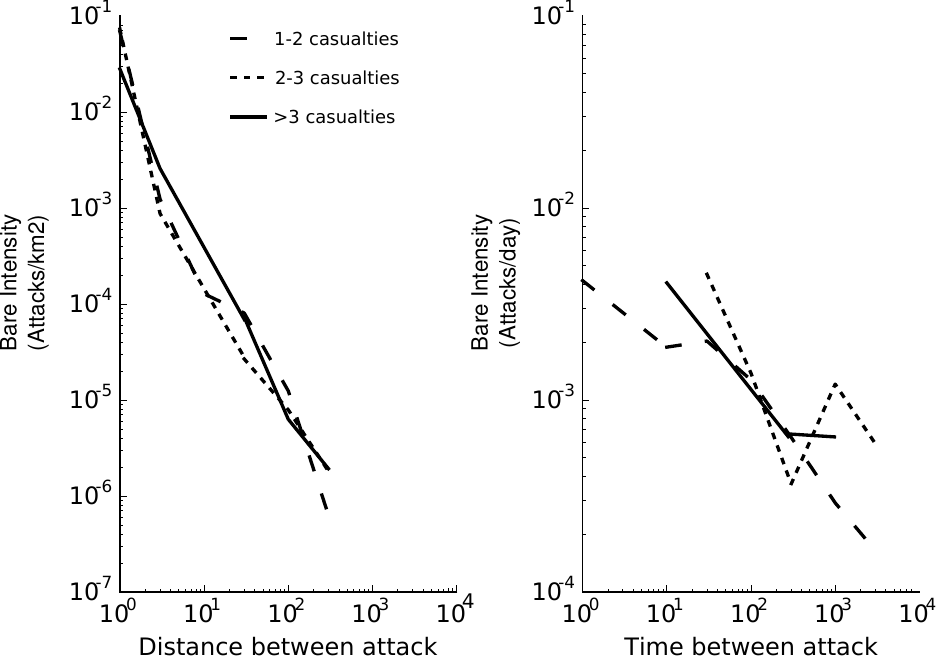}
  \caption{
    Kernel function of the Hawkes process by magnitude
    of attack. Intensity of attacks with respect to
    (left) distance between attacks and (right) time
    between attacks.
  }
  \label{fig:S_magnitudeSparse}
\end{figure*}
        
\begin{figure*}[ht!]
  \centering
  \includegraphics[width=0.9\textwidth]{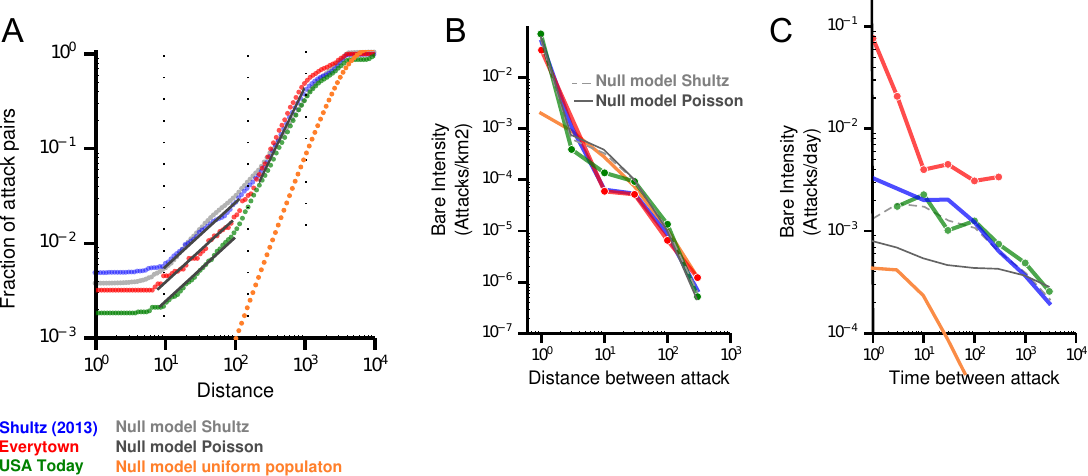}
  \caption{  
    (A) Fraction of attacks within a distance of each other, two null
    models where the attacks are drawn with probabily equal to the
    underlying US population and at times equal to the Schulz database
    (null model Shultz) or with frequency following a Poisson process
    (null model Poisson), and another null model where the attacks are
    drawn from the US area at random and at times equal to the Schulz
    database.
    (B) Intensity of attacks with respect to distance between attacks.  
    (C) Intensity of attacks with respect to time between attacks.
  }
  \label{fig:S_Hawkes}
\end{figure*}

\begin{figure*}[ht!]
  \centering
  \includegraphics[width=0.9\textwidth]{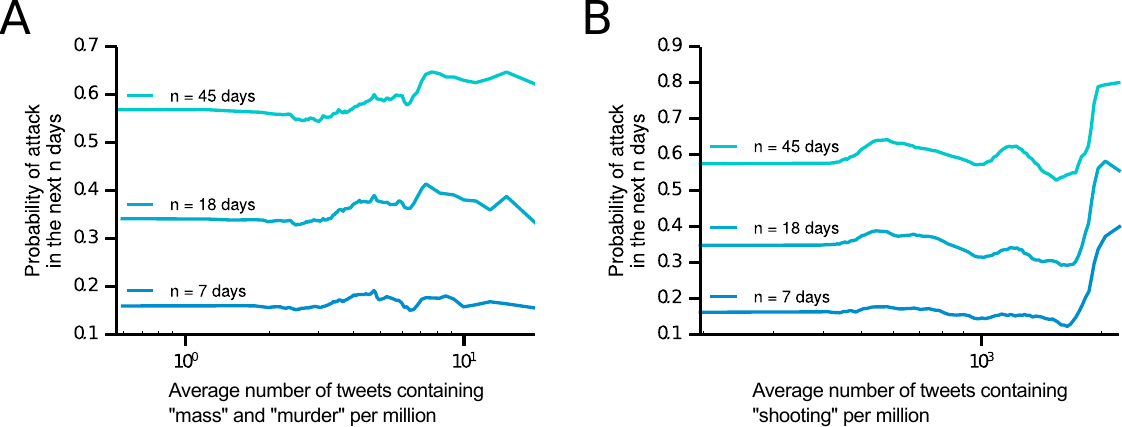}
  \caption{
    Probability of an attack happening in the 7, 18, or 45 days
    following attack $n$, as a function of the mean number of tweets
    with the words (A) ``mass'' and ``murder'' and (B) ``shooting" at
    days $n$ and ${n+1}$. The Shultz database was used for all plots.
  }
  \label{fig:S_copycat_otherWords}
\end{figure*}

\begin{figure*}[ht!]
  \centering
  \includegraphics[width=0.9\textwidth]{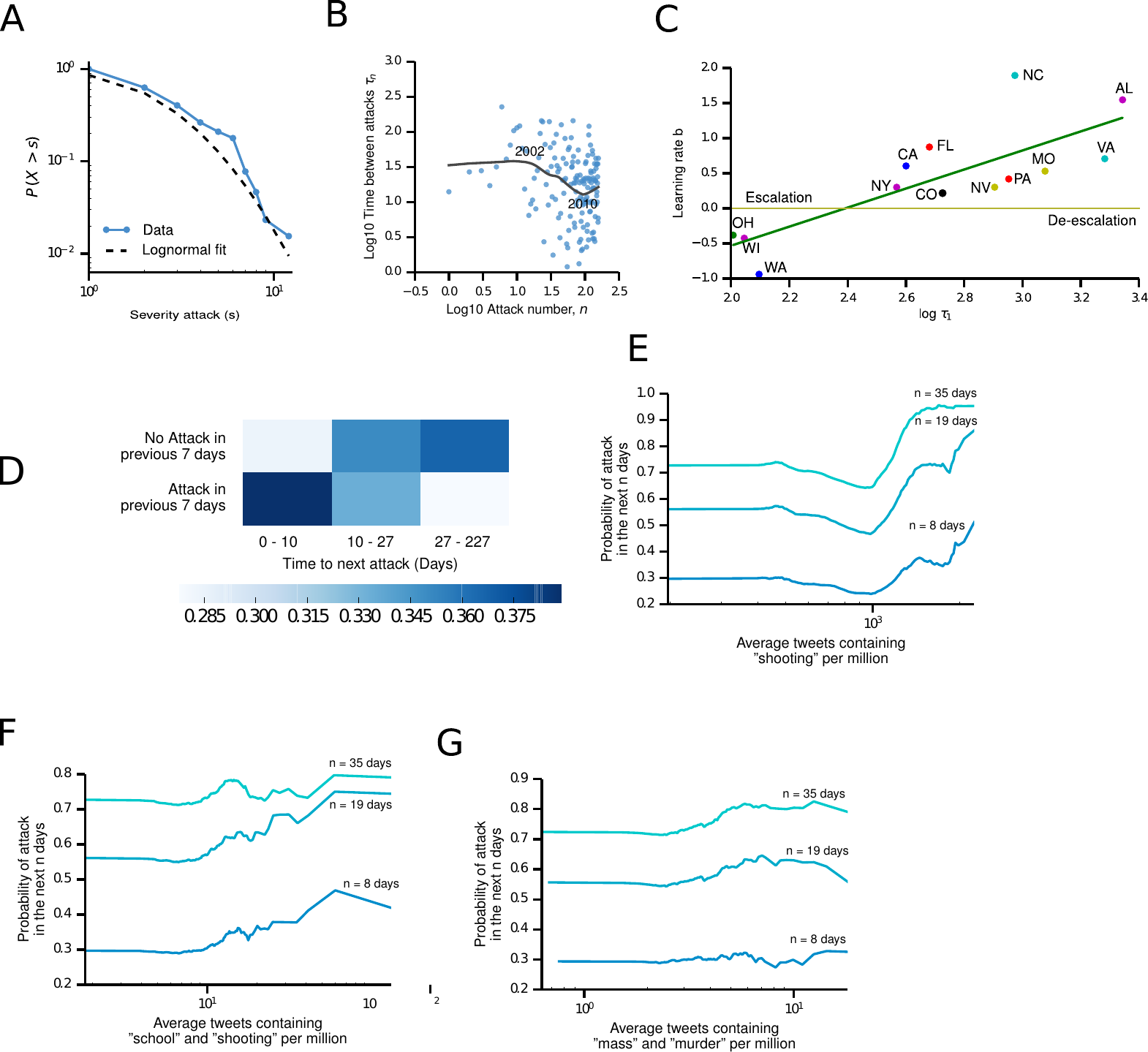}
  \caption{
    \textbf{Active shooting a} 
    (A) Complementary Cumulative Distribution Function (CCDF) for
    event severity (blue dots and solid line) and best fit (dashed
    line) to lognormal distribution.
    (B) The progress curve,
    $\log_{10}{n}$ vs. $\log_{10}{\tau_n}$, for all attacks. LOWESS
    fit ($\delta = 0$, $\alpha = 0.66$) is shown in dark gray, with
    the years where the trend changes annotated.
    (C) Prediction plot,
    $\log_{10}{\tau_1}$ vs. $b$. All states with more than four events
    are considered. States above the $b=0$ line experienced an
    escalation in the number of attacks.
    (D) Probability of attack
    depending on the presence of an attack in the previous seven
    days. Every bin contains one third of the attacks.
    (E--G) Probability of an attack happening in the 8, 19 or 35
    days following to attack $n$, as a function of the mean
    number of tweets talking about shootings at days $n$ and
    ${n+1}$.
  }
  \label{fig:S_FBIa}
\end{figure*}

\begin{figure*}[ht!]
  \centering
  \includegraphics[width=0.9\textwidth]{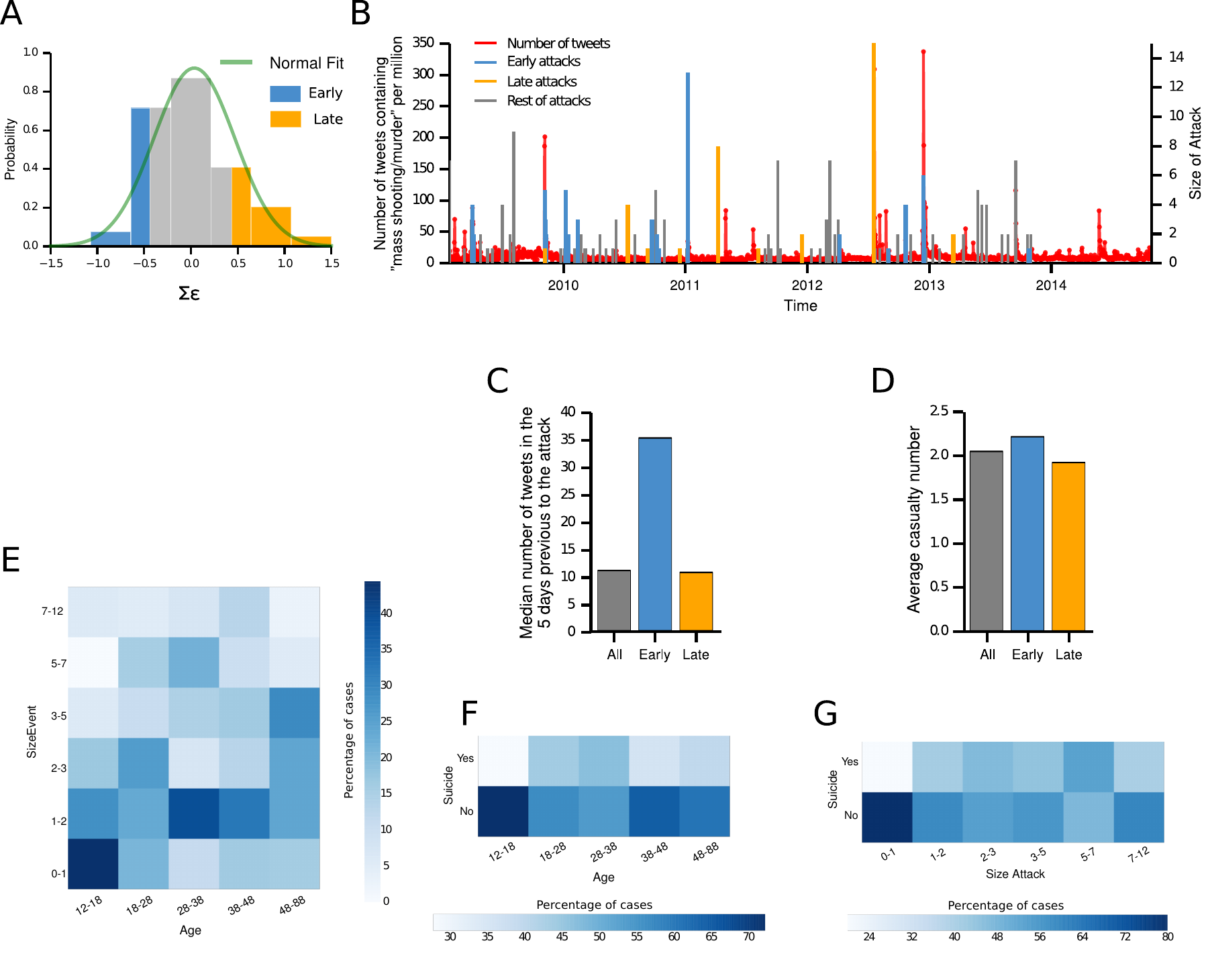}
  \caption{
    \textbf{Active shooting b}  
    (A) Histogram showing $\sum{\epsilon_j}$. \textit{Early} and
    \textit{Late} attacks are marked in blue and orange, respectively.
    (B) Time Series of the number of tweets containing ``mass'' and
    ``shooting'' or ``murder'' (Red lines, left axis), and the size of
    attacks (right axis) for \textit{Early} attacks (Blue),
    \textit{Late} attacks (Orange) and the rest (Grey).
    (C) Median number of tweets in the five days preceding All (Grey),
    \textit{Early} (Blue) and \textit{Late} (Orange) attacks.
    (D) Average casualty number for All (Grey), \textit{Early} (Blue) and
    \textit{Late} (Orange) attacks.
    (E) Probability of different magnitude of events by age group.
    (F) Probability of suicide by age group.
    (G) Probability of suicide by size of attack group.
  }
  \label{fig:S_FBIb}
\end{figure*}

\end{document}